\documentclass[11pt]{article}
\usepackage{latexsym,  amsfonts, amssymb, graphicx, a4, epic, eepic, epsfig,
setspace, lscape}

\usepackage{color}

\def\inst#1{$^{#1}$}

\newtheorem{theorem}{Theorem}[section]
\newtheorem{lemma}[theorem]{Lemma}
\newtheorem{proposition}[theorem]{Proposition}

\newtheorem{definition}[theorem]{Definition}
\newtheorem{corollary}[theorem]{Corollary}
\newtheorem{remark}[theorem]{Remark}

\newcommand{\cX}{{\cal X}}

\newcommand{\cvd}{\hfill\raisebox{3pt}
{\fbox{\rule{0mm}{1mm}\hspace*{1mm}\rule{0mm}{1mm}}\,}\vspace{8pt}}

\def \G {{\Gamma}}

\def \e {{\varepsilon}}
\def \D {{\Delta}}

\def \m {{\mu}}

\def \s {{\sigma}}

\def \g {{\gamma}}
\def \t {{\tau}}

\def \d {{\delta}}
\def \p {{\pi}}

\def\ra{\rightarrow}
\def\Proof{{\sl Proof.}\quad}

\newcommand{\be}[1]{\begin{equation}\label{#1}}
\newcommand{\ee}{\end{equation}}

\newcommand{\bl}[1]{\begin{lemma}\label{#1}}
\newcommand{\el}{\end{lemma}}

\newcommand{\br}[1]{\begin{remark}\label{#1}}
\newcommand{\er}{\end{remark}}

\newcommand{\bt}[1]{\begin{theorem}\label{#1}}
\newcommand{\et}{\end{theorem}}

\newcommand{\bd}[1]{\begin{definition}\label{#1}}
\newcommand{\ed}{\end{definition}}

\newcommand{\bcl}[1]{\begin{claim}\label{#1}}
\newcommand{\ecl}{\end{claim}}

\newcommand{\bp}[1]{\begin{proposition}\label{#1}}
\newcommand{\ep}{\end{proposition}}

\newcommand{\bc}[1]{\begin{corollary}\label{#1}}
\newcommand{\ec}{\end{corollary}}

\newcommand{\bi}{\begin{itemize}}
\newcommand{\ei}{\end{itemize}}

\newcommand{\ben}{\begin{enumerate}}
\newcommand{\een}{\end{enumerate}}

\def \qed {{\square\hfill}}
\def \Z {{\mathbb Z}}
\def \R {{\mathbb R}}

\begin {document}

\title{Sampling from a Gibbs measure with pair interaction by means of PCA}

\author{
Paolo Dai Pra\inst{1}\and
Benedetto Scoppola\inst{2} \and
Elisabetta Scoppola\inst{3}}

%\date{}

\maketitle

\begin{center}
{\footnotesize
\vspace{0.3cm} \inst{1}  Dipartimento di Matematica Pura ed Applicata,
University of Padova\\
Via Trieste 63, 35121 Padova, Italy\\
\texttt{daipra@math.unipd.it}\\

\vspace{0.3cm}\inst{2}  Dipartimento di Matematica, University of Roma
``Tor Vergata''\\
Via della Ricerca Scientifica - 00133 Roma, Italy\\
\texttt{scoppola@mat.uniroma2.it}\\

\vspace{0.3cm} \inst{3} Dipartimento di Matematica, University of Roma
``Roma Tre''\\
Largo San Murialdo, 1 - 00146 Roma, Italy\\
\texttt{scoppola@mat.uniroma3.it}\\ }

\end{center}

\begin{abstract}
We consider the problem of approximate sampling from the finite volume Gibbs measure with a general pair interaction. We exhibit a parallel dynamics (Probabilistic Cellular Automaton) which efficiently implements the sampling. In this dynamics the product measure that gives the new configuration 
in each site
contains a term that tends to favour the original value of each spin. 
This is the main ingredient that  allows to prove that the stationary distribution of the PCA
is close in total variation to the Gibbs measure. 
The presence of the parameter that drives 
the "inertial" term mentioned 
above gives the possibility to control the degree
of parallelism of the numerical implementation of the dynamics.
\end{abstract}

\eject

\tableofcontents
%%%%%%%%%%%%%SECT 1%%%%%%%%%%%%%%%%%%%%%%%%%
\section{Introduction}
\label{Intro}
\vspace{0.3cm}

Probabilistic Cellular Automata (PCA) are (time-homogeneous) discrete-time Markov Chains on a product space $S^V$, whose transition probability $P(d\s|\s')$ is a product measure:
\[
P(\s|\s') = \prod_{i \in V} p_i(d\s_i|\s'),
\]
where, for $i \in V$ and $\s' \in S^V$, $p_i(d\s_i|\s')$ is a probability on $S$. Compared with the more familiar {\em sequential} dynamics, where the transition probabilities $P(d\s|\s')$ are supported on {\em configuration} $\s$ with $\s_j = \s'_j$ for all but one $j \in V$, PCA's exhibit the following peculiar features.

\bi
\item
The {\em parallel updating} rule allow to exploit the efficiency of {\em parallel computation} in the simulation of these dynamics, making them desirable Markov Chain Monte Carlo algorithms.
\item
PCA's give rise to well defined {\em infinite volume} dynamics ($V$ infinite countable), without passing to continuous time.
\ei
The study of PCA's in the context of Equilibrium Statistical Mechanics dates back to \cite{GKLM,LMS}, where various features of the infinite-volume limit have been investigated, in particular its space-time Gibbsian nature. On the other hand, invariant measures for infinite-volume PCA's may be non-Gibbsian, as shown in \cite{FT}.

In the context of Markov Chain Monte Carlo algorithms the following natural problem arise: given a probability $\mu$ on $S^V$, construct a PCA whose invariant measure is $\mu$; in particular, in the case $\mu$ is a Gibbs measure for a short range interaction, one expects that the transition probabilities of the PCA can be chosen to be {\em local}, i.e. $p_i(d\s_i|\s')$ depends only on $\s'_j$ for $j$ ``close'' to $i$. While Markov Chain with sequential dynamics having these features can always be constructed, the existence of a PCA with the given invariant measure $\mu$ is not granted. Counterexamples are given in \cite{D}, while \cite{KV} provides explicit conditions on $\mu$ for the existence of a PCA {\em reversible} with respect to $\mu$.

 A well understood example is that of the $2d$ Ising model. By the results in \cite{KV} it follows that no
 PCA can be reversible with respect to the $2d$ Ising model. In \cite{LMS} and \cite{CN} a PCA is introduced whose invariant (reversible) measure $\pi$
 is related to the Ising model as follows: the projection of $\pi$ to the {\em even} sites, i.e. those $(i,j) \in \Z^2$ with $i+j$ even, coincides with the same projection 
 of the Ising model, and the same holds for odd sites; however, under $\pi$, spins at even sites are independent from spins at odd sites, unlike for the Ising model. 
 
 When the
 nearest neighbor interaction of the Ising model is generalized to a general pair interaction, this simple structure is lost.
In this paper, following the ideas introduced in \cite{ISS}, we present a simple  way to modify and extend the PCA in \cite{CN}, and use it to sample approximately from a Gibbs measure.
Given a spin configuration $\s \in \{-1,1\}^V$, where $V$ is a finite subset of $\Z^d$, we start with a Hamiltonian of the form
\[
H(\s) := -\sum_{i,j} J_{ij} \s_i \s_j,
\]
corresponding to the Gibbs measure
\[
\pi^G(\s) \propto \exp[-H(\s)].
\]
This Hamiltonian can be {\em lifted} to a Hamiltonian on $ \left(\{-1,1\}^V\right)^2$, setting
\[
H(\s,\s') :=  -\sum_{i,j} J_{ij} \s_i \s'_j + q\sum_i (1- \s_i \s'_i).
\]
The measure
\[
\mu_2(\s,\s') \propto \exp[-H(\s,\s')]
\]
is such that the conditional measure
\[
\mu_2(\s|\s') = \frac{\mu_2(\s,\s')}{\sum_{\tau} \mu_2(\tau,\s')}
\]
is a product measure, and can therefore be taken as transition probability of a PCA, which turns out to be reversible for a probability $\pi^{PCA}$.
The parameter $q$  controls 
 the average number of spin-flips in a single step of our dynamics. This is the analogous
 of the self-interaction 
  considered in \cite{CNS}  to study metastability in the limit of zero temperature,
 but our regime and our goal are  completely different.
 The parameter $q$ acts as the brake of the dynamics: for large values of $q$ the dynamics
is very slow, flipping few spins at each time, tending to "freeze" the system in its  configuration,
while a dynamics with $q=0$ is  for instance  the case of \cite{CN}.
We want to show that
for suitable choices of $q$ we have a dynamics that is considerably faster than the
usual single spin-flip dynamics, and tends to a stationary measure $\pi^{PCA}$ that can be shown
to be very close to the Gibbs measure $\pi^G$.

% ****(paolo per altre referenze )****

More precisely we prove in Theorem \ref{teo} that when the volume goes to infinity, $|V|\to\infty$,
the total variation distance between  $\pi^{PCA}$ and $\pi^G$ goes to zero when $q$ is such that
the mean density of flipped spins, $\d :=e^{-2q}$ satisfies $\lim_{|V|\to\infty}\d^2|V|=0$.
Note that this request is compatible with a choice of $q$ such that the average number of spin-flips in a move, $\d|V|$, is large.
In Section \ref{proof} we give the proof of this convergence for rather general two-body interactions. Given the generality of the model, convergence is proved under Dobrushin uniqueness conditions, which implies the minimal condition needed for our argument, namely a form of fast decay of correlations. For special models, such as ferromagnets in pure states (see \cite{N}), we expect that this fast decay of correlations hold true also in the coexistence region, where Dobrushin uniqueness fails.

%As far as the discussion on the ``gain" obtained by updating many spin in a single move, this depends on
%the particular interaction considered. We give in Section \ref{CW} the simple example of the
%mean-field Ising model, where the phase diagram of the PCA is compared with the kwnon results
%on the model and the convergence to equilibrium of the PCA is compared with the convergence
%of single spin-flip dynamics, as discussed for instance in \cite{LLP}. The gain is definite, however
%we have to note that this is a case where mixing is fast, and so is not a good example
%to compare the performances of PCA vs single spin flip dynamics.
%Certainly the parallel dynamics proposed in this paper  for sampling from a Gibbs measure can be effectively exploited  in simulations performed with parallel calculus. 
%However the discussion of a real gain of the parallel dynamics is not given in the present paper 
%but a  more detailed discussion
%on the mixing time for PCA will be the subject of a future paper. 
%

%%%%%%%%%%%%%

\subsection{Definitions} 
\label{def}
Given a finite volume $V\subset\Z^d$ we consider the spin configurations
$\s\in\{-1,1\}^V$ and  define
\be{ham}
H(\s)=-\sum_{i\ne j}J_{ij}\s_i\s_j
\ee
where $(J_{ij})_{i,j \in Z^d}$ is a given infinite {\em symmetric} matrix satisfying
\be{J}
\sup_i\sum_j|J_{ij}|=J<\infty
\ee
For simplicity, the sum in (\ref{ham}) is supposed to range over $i,j \in V$; in other words, boundary conditions are empty. More general boundary conditions could be treated with no
difficulty. \\
The Hamiltonian will be written equivalently
\be{ham2}
H(\s)=\sum_{i\in V}h_i(\s)\s_i
\ee
where 
\be{hi}
h_i(\s)= -\sum_jJ_{ij}\s_j
\ee
Then we can define the standard Gibbs measure as
\be{gibbs}
\pi^G(\s)=\frac{e^{-H(\s)}}{Z^G}\equiv\frac{w^G(\s)}{\sum_{\s}w^G(\s)}
\ee
where
\be{zg}
Z^G=\sum_{\s}e^{-H(\s)}, \qquad w^G(\s)=e^{-H(\s)}
\ee
A sampler from this Gibbs measure can be realized by defining a
Markov chain defined on the state space
$\{-1,1\}^V$ with invariant measure $\pi^G$, following the usual ideas
of Markov Chain Monte Carlo methods. A standard algorithm is the {\it Gibbs sampler}
which at each integer time: 
\bi
\item[-] a site $i\in V$ is randomly chosen (with uniform distribution);
\item[-] the configuration outside $i$ is left unchanged;
\item[-] the new spin at $i$ is sampled from the conditional measure $\p^G(\, \cdot \, | \s_{V \setminus \{i\}})$
\ei
This is equivalent to define:
\be{gla}
P^{G}_{\s,\t}
=\cases{\frac{1}{|V|}{e^{h_i(\s)\s_i}\over e^{h_i(\s)\s_i}+e^{-h_i(\s)\s_i}}&if $\t=\s^i$\cr
1-\sum_{i \in V}P^G_{\s,\s^i}&if $\s=\t$\cr 0&otherwise\cr}
\ee

Different single spin flip dynamics can also be defined,  for instance with Metropolis rates.
%It is well known that $\pi^G$ is the stationary measure of the Markov chain
%known as {\it Glauber dynamics}, i.e, a Markov chain defined on the 
%configurations $\s\in\{-1,1\}^V$ by the following transition probabilities
%\be{gla}
%P_{\s,\s'}=\cases{\frac{1}{|V|}e^{-\max\{0,H(\s')-H(\s)\}}&if $d(\s,\s')=\frac{1}{2}\sum_i|\s_i-\s'_i|=1$\cr
%0&if $d(\s,\s')>1$\cr
%1-\sum_{\t\ne\s}P_{\s,\t}&if $\s=\s'$\cr}
%\ee
On the other side we can define an alternative collective dynamics, that we will call
{\it PCA dynamics}, in the following way.
Define
\be{hampca}
H(\s,\s')=\sum_{i\in V}\Big[h_i(\s)\s'_i+q(1-\s_i\s'_i)\Big]
\ee
where $q>0$.
The PCA dynamics is the Markov chain defined by the following 
transition probabilities
\be{pss}
P^{PCA}_{\s,\s'}=\frac{e^{-H(\s,\s')}}{Z_\s}
\ee
where
\be{zs}
Z_\s=\sum_\t e^{-H(\s,\t)}=w^{PCA}(\s)
\ee
It is a standard task to show that the chain is reversible with respect to the measure
\be{ps}
\pi^{PCA}(\s)=\frac{\sum_\t e^{-H(\s,\t)}}{\sum_{\t,\t'} e^{-H(\t,\t')}}\equiv
\frac{w^{PCA}(\s)}{\sum_{\t}w^{PCA}(\t)}=\frac{w^{PCA}(\s)}{Z^{PCA}}
\ee

Note that $\pi^{PCA}$ turns out to be the marginal of the Gibbs measure on 
the space of pairs of configurations
\be{mu2}
\m_2(\s,\t)={ e^{-H(\s,\t)}\over {Z^{PCA}}}.
\ee

Due to the definition (\ref{hampca}) the transition probabilities of this Markov chain
can be written as a product of the transition probability of each component  $\s'_i$ of the new configuration,
as usual for PCAs:
$$
P^{PCA}_{\s,\s'}=\prod_{i\in V}P(\s'_i|\s),
$$
where
\[
P(\s'_i|\s) = \frac{\exp[\s'_i (h_i(\s) - q \s_i)]}{2 \cosh(h_i(\s) - q \s_i)}.
\]
Our goal is to show that, for a suitable choice of ``moderately'' large $q$:
\begin{itemize}
\item[-]
the invariant measure of the PCA $\pi^{PCA}$ is ``close'' to $\pi^G$;
\item[-]
the PCA updates at each time step a large number of spins.
\end{itemize}

In order to state precisely the results we obtain, we recall  the total variation distance, or $L_1$ distance, 
between $\pi^G$ and $\pi^{PCA}$ as
\be{dist}
\| \pi^{PCA}-\pi^{G}\|_{TV}=\frac{1}{2}\sum_\s|\pi^{PCA}(\s)-\pi^{G}(\s)|
\ee

%%%%%%%%%%%%%%%%%%%%%%%%%%%
\subsection{Results} 
\label{result}

Before stating our main result, we make some remarks on the relation between $\pi^G$ and $\pi^{PCA}$. Note first of all that 
$$
w^{PCA}(\s)=\sum_\t e^{-\sum_{i\in V}[h_i(\s)\t_i+q(1-\s_i\t_i)]}=
\sum_{I\subset V}e^{-\sum_{i\in V}h_i(\s)\s_i+2\sum_{i\in I}h_i(\s)\s_i-2q|I|}=$$
\be{ww}
=w^{G}(\s)\prod_{i\in V}(1+\delta\phi_i)
\ee
where $\d=e^{-2q}$ and 
$$\phi_i=e^{-2\sum_jJ_{ij}\sigma_i\sigma_j}$$
We will call
\be{f}
f(\sigma)=\prod_{i\in V}(1+\delta\phi_i).
\ee
It easily follows that
\be{pgppca}
\p^{PCA}(\s)=\p^{G}(\s)\frac{f}{\p^G(f)}
\ee
We also define the probability $\tilde{\pi}$ by
\be{ptilde}
\tilde\p(\s)=\p^{G}(\s)\frac{f^2}{\p^G(f^2)}
\ee
\bt{teo1}
For any $q\ge 0$ let 
$\d:=e^{-2q}$. Suppose:
\begin{itemize}
\item[(a)]
$\d = \d(|V|)$ is such that $\lim_{|V|\to\infty}\d^2|V|=0$;
\item[(b)]
there exists $\d_0 > 0$ such that
\be{condvar}
\sup_V\sup_{\d\in[0,\d_0]}\frac{1}{|V|}Var_{\pi}\left[\sum_{i\in V} \frac{\phi_i}{1+\d\phi_i}\right]<\infty 
\ee
for $\pi = \pi^{PCA}$ and $\pi = \tilde{\pi}$.
\end{itemize}
Then
\be{conv}
\lim_{|V|\to\infty}\| \pi^{PCA}-\pi^{G}\|_{TV}=0.
\ee
\et
Condition( \ref{conv}) follows by controlling the decay of correlations of the family of functions $\{\phi_i\}$. This control can be achieved by using the {\em Dobrishin uniqueness condition}. More precisely, we obtain the following result.
\bp{teo}
Assume
\be{assumption}
\sup_i \sum_{j} \tanh(2 \left|J_{i,j}\right|) < 1.
\end{equation}
Then assumption (b) in Theorem \ref{teo1} holds.
\ep

\subsection{Discussion and open problems}

\bi
\item[1-]
{\bf Low temperature}

Hypothesis (\ref{assumption}) is the Dobrushin condition for uniqueness of phase  for the 
Gibbs measure $\p^G$. It is needed due to the generality of the interaction we are considering.
Since the crucial ingredient in the proof of the Theorem is the correlation decay, we expect that hypothesis (\ref{assumption}) can be weakened if additional assumptions
on the interaction are considered. This will be discussed in a forthcoming paper.

 In the last part of this paper we discuss in detail the Curie Weiss model. In the uniqueness region we show that (\ref{conv}) hold with the condition $\lim_{|V| \to \infty} \d = 0$, much weaker than condition (a) in Theorem \ref{teo1}, while $\lim_{|V| \to \infty} \d^2 |V|= 0$ suffices also at low temperature for a modified dynamics which is forced to select configurations with positive magnetization.

%%%%%%%%%%%%%%%%

\item[2-]
{\bf Convergence to equilibrium of PCA}

Note that when $\d=\frac{1}{|V|}$ the PCA dynamics is essentially equivalent to sequential Gibbs sampler, since the average number of spins that are updated in a time step is of order $\d |V|$.
A natural  question is then to compare the speed of convergence to equilibrium of the PCA dynamics
for $\frac{1}{|V|}\ll \d\ll \frac{1}{\sqrt{|V|}}$
vs single spin flip dynamics. 
This is of course a central problem in applications.
A quantitative comparison of the two dynamics is beyond the purposes of this paper. Our aim is rather to understand to what extent sampling of Gibbs measures can be implemented by parallel dynamics.

%%%%%%%%%%%%%%%%
\item[3-]
{\bf Applications}

The PCA dynamics discussed  in this paper has been introduced in \cite{ISS} in order to study
the clique problem on large graphs. In that case the general  setup was  more complicated since the canonical ensemble was considered. The excellent numerical results
obtained for the clique problem, encouraged us to better undersatnd the PCA dynamics. In \cite{GSSV}
a phase transition in the case of random graphs was proved. In general the extension of the result of this paper to the canonical ensemble is an interesting problem, and is currently under investigation.

%%%%%%%%%%%%%%%

%%%%%%%%%%%%%%%%%%%%%%%%

\ei

%%%%%%%%%%%%%%%%%%%Sec2%%%%%%%%%%%%%
%%%%%%%%%%%%%%%%%%%%%%%%%%%%%%%%%%%
\section{Proofs}
\label{proof}

\subsection{Proof of Theorem \ref{teo1}} 

Using (\ref{ww}), (\ref{f}) and (\ref{pgppca}), we have
$$
\| \pi^{PCA}-\pi^{G}\|_{TV}=\sum_\s \frac{w^{G}(\s)}{Z^G}\left| \frac{\pi^{PCA}(\s)}{\pi^{G}(\s)}-1\right|
=\sum_\s \frac{w^{G}(\s)}{Z^G}\left| \frac{w^{PCA}(\s)}{w^{G}(\s)}\,\frac{Z^{G}}{Z^{PCA}}-1\right|
$$
\be{stid}
=\sum_\s \frac{w^{G}(\s)}{Z^G}\left| \frac{f(\s)}{\pi^{G}(f)}-1\right|=
\pi^{G}\left(\left| \frac{f(\s)}{\pi^{G}(f)}-1\right|\right)\le\frac{(\rm{var}_{\pi_G}(f))^{1/2}}{\pi_G(f)}
\ee
Therefore we need an estimate on the dependence on $\d$ of the quantity
\be{Del}
\D(\d)=\frac{\pi_G(f^2)}{(\pi_G(f))^2}-1
\ee
More precisely, we want to show that
\be{uno}
{1\over |V|}\ln \pi^G(f^2)-{2\over |V|}\ln \pi^G(f)=O(\delta^2)
\ee
Note first that, writing
\be{flog}
f(\s)=\exp\left[\sum_{i\in V}\log(1+\d\phi_i(\s))\right]
\ee
we have
\be{duno}
\frac{d}{d\d}\log \p^G[f]=\p^{PCA}\left[\sum_{i\in V} \frac{\phi_i}{1+\d\phi_i}\right]
\ee
and
\be{ddue}
\frac{d^2}{d^2\d}\log \p^G[f]=-\p^{PCA}\left[\sum_{i\in V} \left(\frac{\phi_i}{1+\d\phi_i}\right)^2\right]
+Var_{\p^{PCA}}\left[\sum_{i\in V} \frac{\phi_i}{1+\d\phi_i}\right],
\ee
where we have used (\ref{pgppca}).
Analogously, using (\ref{ptilde}),
we have
\be{dquno}
\frac{d}{d\d}\log \p^G[f^2]=2\tilde\p\left[\sum_{i\in V} \frac{\phi_i}{1+\d\phi_i}\right]
\ee
and
\be{dqdue}
\frac{d^2}{d^2\d}\log \p^G[f^2]=-2\tilde\p\left[\sum_{i\in V} \left(\frac{\phi_i}{1+\d\phi_i}\right)^2\right]
+4Var_{\tilde\p}\left[\sum_{i\in V} \frac{\phi_i}{1+\d\phi_i}\right].
\ee
The idea is to exploit this explicit results in order to control 
up to the second order an expansion of (\ref{uno}) around $\d=0$.
Clearly the first order computed in $\d=0$ exhibit an explicit 
cancellation, since for $\d=0$ we have that $\p^G=\p^{PCA}=\tilde\p$.
In order to show (\ref{uno}), therefore, it is enough to prove that
\[
\sup_V\sup_{\d\in[0,\d_0]}\frac{1}{|V|} \left(\left|\frac{d^2}{d^2\d}\log \p^G[f]\right| + \left| \frac{d^2}{d^2\d}\log \p^G[f^2] \right| \right) < +\infty,
\]
which, by (\ref{ddue}) and (\ref{dqdue}), follows from (\ref{condvar}).

$\qed$

%%%%%%

\subsection{F\"{o}llmer's estimate}

Let $\pi$ be a probability of $\{-1,1\}^{\Z^d}$, and denote by $\pi\left(\s_i | \s_{\backslash i}\right)$ its local specifications. Define, for $i \neq j$, the {\em Dobrushin coefficients}:
\[
\g_{ij} = \sup_{\s} \left| \pi\left(\s_i =1 | \s_{\backslash i}\right)-  \pi\left(\s_i =1 | \s^j_{\backslash i}\right) \right|.
\]
Assume the so-called {\em Dobrushin uniqueness condition}:
\be{dobuniq}
\g := \sup_i \sum_j \g_{ij} < 1.
\ee
Denoting by $\Gamma$ the matrix with elements $\g_{ij}$, under (\ref{dobuniq}) the matrix
\[
D := \sum_{n=0}^{+\infty} \G^n
\]
is well defined.
For a function $f: {-1,1}^{\Z^d} \ra \R$, set 
\[
\rho_j(f) := \sup_{\s} \left| f(\s) - f(\s^j) \right|.
\]
The following is the main result of the beautiful paper \cite{Fo82} by H. F\"{o}llmer  (se also \cite{Ku82} for related results).

\bt{fo}
\be{cov}
\left|Cov_{\pi}(f,g) \right| \leq \frac{1}{4} \sum_{i,j} D_{ij} \rho_i(f) \rho_j(g).
\ee
\et
%%%%%%

\subsection{Proof of Proposition \ref{teo}}

%Note that $\varphi_i = \varphi_0 \circ \theta_i$. Moreover, using the facts that $\varphi_0 \geq 0$ and that the map $x \mapsto \frac{x}{1+\d x}$ has Lipschitz constant $1$ on $[0,+\infty)$, we have that
%\[
%\rho\left(\frac{\varphi_0}{1+ \d \varphi_0} \right) \leq \rho(\varphi_0).
%\]
%Thus, using Theorem \ref{ku} and Remark \ref{rku}, estimate (\ref{uno}) follows once we show that (\ref{cond1}) and (\ref{cond2}) hold for $\pi = \pi^{PCA}$ and $f = \varphi_0$, uniformly for $\d$ small enough. We show it under the following assumption:
%\begin{equation}
%\label{assumption1}
%\sum_{i} \tanh(2 \left|J_{0,i}\right|) < 1 \hspace{1cm} \sum_i \left|J_{0,i}\right| |i|^{\e} < +\infty
%\end{equation}
%for some $\e>0$. 
%This is the Dobrushin condition for uniqueness of phase  for the 
%Gibbs measure $\p^G$ with an additional condition on summability of intaractions. 

We begin by showing that, under (\ref{assumption}), the Dobrushin Uniqueness condition hold for $\pi^{PCA}$ and $\tilde{\pi}$, for $\d$ sufficiently small.
 
\bp{dob1}
Let $\g_{i,j}$ be the Dobrushin coefficients for $\pi^{PCA}$ (resp. $\tilde{\pi}$). Then
\[
\g_{i,j} \leq \tanh(2 \left|J_{i,j}\right|) + \frac{1}{2}\rho_j(\psi_{i,\d}),
\]
where $\psi_{i,\d}$ is defined by
\[
2\psi_{i,\d}(\s) = \log \frac{1+\d e^{-2 h_i(\s)}}{1+\d e^{2 h_i(\s)}} + \sum_{l \neq 0} \log \frac{1+\d \exp[-2 J_{i,l} \s_l - 2 \s_l h_{i,l}(\s)]}{1+\d \exp[2 J_{i,l} \s_l - 2 \s_l h_{i,l}(\s)]}
\]
with
\[
h_{i,l}(\s) = -\sum_{j \neq l} J_{i,j} \s_j.
\]
\ep

\Proof
The proof consists in a rather direct and straightforward computation. We give the proof for $\pi = \pi^{PCA}$. The proof for $\tilde{\pi}$ is similar. Set, for simplicity, $i = 0$.
We write
\[
H_i := \log(1+\d \phi_i),
\]
so that
\[
\pi^{PCA}(\s) = \frac{1}{Z^{PCA}} \exp\left[  \sum_{i,j} J_{i,j} \s_i \s_j  + \sum_i H_i(\s)\right].
\]
Note that
\[
 \sum_{i,j} J_{i,j} \s_i \s_j =- 2 \s_0 h_0(\s) + C_1(\s_{\backslash 0}),
 \]
 where, with $C_1(\s_{\backslash 0})$ we denote all remaining terms which do not depend on $\s_0$. Similarly
 \[
 H_0(\s) = \log \left(1+\d e^{-2 \s_0 h_0(\s)} \right) = \frac{1}{2} \s_0 \log \frac{1+\d e^{-2 h_0(\s)}}{1+\d e^{2 h_0(\s)}} + C_2(\s_{\backslash 0}),
 \]
 and, for $l \neq 0$,
 %\begin{multline*}
 $$
 H_l(\s) = \log \left(1+\d e^{-2 \s_l \sum_j J_{l,j} \s_j} \right) = \log \left( 1+\d \exp[-2 J_{0,l} \s_0 \s_l - 2  \s_l h_{0,l}(\s)]\right)  $$
 $$= \frac{1}{2} \s_0 \log \frac{\left( 1+\d \exp[-2 J_{0,l}  \s_l - 2  \s_l h_{0,l}(\s)]\right)}{\left( 1+\d \exp[2 J_{0,l}  \s_l - 2  \s_l h_{0,l}(\s)]\right)} + C_3(\s_{\backslash 0}).
$$
% \end{multline*}
 It follows that
 \[
 \pi^{PCA}(\s_0=1|\s_{\backslash 0}) = \frac{\exp\left[ - 2 h_0(\s) + \psi_{\d}(\s)\right]}{2 \cosh (2 h_0(\s) - \psi_{\d}(\s))}.
 \]
 Now, writing $\psi_{\d}$ for $\psi_{0,\d}$, we have
 \[
 - 2 h_0(\s) + \psi_{\d}(\s) - \left(- 2 h_0(\s^j) + \psi_{\d}(\s^j)\right) = 4 J_{0,j} + \psi_{\d}(\s) -  \psi_{\d}(\s^j).
 \]
 Setting $x :=   2 h_0(\s) - \psi_{\d}(\s)$ and $y := 4 J_{0,j} + \psi_{\d}(\s) -  \psi_{\d}(\s^j)$, we have
 \[
 \pi^{PCA}(\s_0=1|\s_{\backslash 0}) - \pi^{PCA}(\s_0=1|\s^j_{\backslash 0}) = \frac{e^{-x}}{2 \cosh(x)} - \frac{e^{-x-y}}{2 \cosh(x+y)} =: g_y(x).
 \]
 Unless $y=0$ (which gives $g_y \equiv 0$), the derivative $\left(g_y^2\right)'$ vanishes only at $x = -y/2$, where $g^2_y$ attains its absolute maximum $\tanh^2(y/2)$. This yields
 \[
 \left|\pi^{PCA}(\s_0=1|\s_{\backslash 0}) - \pi^{PCA}(\s_0=1|\s^j_{\backslash 0}) \right| \leq \tanh\left(2 J_{0,j} + \frac{1}{2}\left|\psi_{\d}(\s) -  \psi_{\d}(\s^j)\right|\right).
 \]
 Since, for every $a,b \geq 0$, $\tanh(a+b) \leq \tanh(a) + b$, the conclusion of the lemma follows.
 \cvd

 To complete the proof that   the Dobrushin Uniqueness condition hold for $\pi^{PCA}$ and $\tilde{\pi}$, for $\d$ sufficiently small, it is enough to show the following result.
 \bl{dob2}
 We have
 \[
 \sup_i\sum_{j}  \rho_j(\psi_{i,\d}) = o(\d).
 \]
 as $\d \rightarrow 0$.
 \el
 \Proof
 Set $i=0$ and $\psi_{0,\d} = \psi_{\d}$. The estimate for a generic $i$ is similar.
 Ignoring an irrelevant factor $2$
% \begin{multline*}
$$
  \rho_j(\psi_{\d})  \leq \rho_j\left(\log \frac{1+\d e^{-2 h_0(\s)}}{1+\d e^{2 h_0(\s)}}\right) + \rho_j \left(\log \frac{1+\d \exp[-2 J_{0,j} \s_j - 2 \s_j h_{0,j}(\s)]}{1+\d \exp[2 J_{0,j} \s_j - 2 \s_j h_{0,j}(\s)]}\right) $$
  $$+  \sum_{l \neq 0,j}\rho_j\left( \log \frac{1+\d \exp[-2 J_{0,l} \s_l - 2 \s_l h_{0,l}(\s)]}{1+\d \exp[2 J_{0,l} \s_l - 2 \s_l h_{0,l}(\s)]}\right).
$$
%  \end{multline*}
  The main difficulty comes from the third term, and we only deal with it, i.e. we show that
  \[
  \sum_{j} \sum_{l \neq 0,j}\rho_j\left( \log \frac{1+\d \exp[-2 J_{0,l} \s_l - 2 \s_l h_{0,l}(\s)]}{1+\d \exp[2 J_{0,l} \s_l - 2 \s_l h_{0,l}(\s)]}\right) = o(\d).
  \]
Set 
\[
C_l(\s) := \log \frac{1+\d \exp[-2 J_{0,l} \s_l - 2 \s_l h_{0,l}(\s)]}{1+\d \exp[2 J_{0,l} \s_l - 2 \s_l h_{0,l}(\s)]} = \log\left( 1-2\d\frac{\sinh(2J_{0,l}\s_l) \exp[-2\s_l h_{0,l}(\s)]}{1+\d \exp[2J_{0,l} \s_l -2\s_l h_{0,l}(\s)]}\right).
\]
It is not restrictive to assume that $\d$ is small so that 
\[
2\d\frac{\sinh(2 \left|J_{0,l}\right|) \exp[-2\s_l h_{0,l}(\s)]}{1+\d \exp[2J_{0,l} \s_l -2\s_l h_{0,l}(\s)]} < \frac{1}{2}.
\]
Since, on $(-1/2,1/2)$ the map $x \mapsto \log(1-x)$ has Lipschitz constant $2$, we have
%\begin{multline*}
$$
\left|C_l(\s) - C_l(\s^j)\right| \leq 4 \d \sinh(2 |J_{0,l}|) \left| \frac{e^{-2\s_l h_{0,l}(\s^j)}}{1+\d e^{2J_{0,l}\s_l - 2\s_l h_{0,l}(\s^j)}} - \frac{e^{-2\s_l h_{0,l}(\s)}}{1+\d e^{2J_{0,l}\s_l - 2\s_l h_{0,l}(\s)}} \right|  $$
$$\leq  4 \d^2 e^{2J}  \sinh(2 |J_{0,l}|) \left| e^{-2\s_l h_{0,l}(\s^j)} - e^{-2\s_l h_{0,l}(\s^)}\right| $$
$$ \leq
 4 \d^2 e^{4J}  \sinh(2 |J_{0,l}|) \left|h_{0,l}(\s^j) - h_{0,l}(\s)\right| \leq 4 \d^2 e^{6J} |J_{0,l}| |J_{j,l}|,
$$
where we have used the estimates $e^{2J_{0,l}\s_l - 2\s_l h_{0,l}(\s^j)} \leq 2J$, $\sinh(2 |J_{0,l}|) \leq e^{2J} |J_{0,l}|$. It follows that
\[
\sum_{j}   \sum_{l \neq 0,j}\rho_j\left( \log \frac{1+\d \exp[-2 J_{0,l} \s_l - 2 \s_l h_{0,l}(\s)]}{1+\d \exp[2 J_{0,l} \s_l - 2 \s_l h_{0,l}(\s)]}\right) \leq 4 \d^2 e^{6J}\left( \sum_j  |J_{0,j}|\right)^2 = o(\d).
\]

\cvd

\noindent
{\em Proof of Proposition \ref{teo}}. For $\pi = \pi^{PCA}$ or $\pi = \tilde{\pi}$ we have, by Theorem \ref{fo},
$$
\frac{1}{|V|}Var_{\pi}\left[\sum_{i\in V} \frac{\phi_i}{1+\d\phi_i}\right] = \frac{1}{|V|}\sum_{i,j \in V} Cov_{\pi} \left(  \frac{\phi_i}{1+\d\phi_i},  \frac{\phi_j}{1+\d\phi_j} \right)  $$ $$ \leq 
 \frac{1}{|V|}\sum_{i,j \in V} \sum_{h,k} D_{hk} \rho_h\left(  \frac{\phi_i}{1+\d\phi_i}\right)\rho_k\left(  \frac{\phi_j}{1+\d\phi_j}\right).
$$
 Since  the map $x \mapsto \frac{x}{1+\d x}$ has Lipschitz constant $1$ on $[0,+\infty)$, we have that
\[
\rho_h\left(\frac{\phi_i}{1+ \d \phi_i} \right) \leq \rho_h(\phi_i).
\]
Moreover, it is easily seen that
\[
\rho_h(\phi_i) \leq e^{2J} |J_{i,h}|.
\]
Therefore, since 
\[
\sup_i \sum_j \left(\G^n\right)_{ij} \leq \g^n
\]
which implies
\[
\sup_h \sum_k |D_{hk}| \leq \frac{1}{1-\g},
\]
we get
$$
\frac{1}{|V|}Var_{\pi}\left[\sum_{i\in V} \frac{\phi_i}{1+\d\phi_i}\right] \leq  e^{4J} \frac{1}{|V|}\sum_{i,j \in V} \sum_{h,k \in V} D_{hk} |J_{i,h}| |J_{j,k}| $$ $$\leq J^2e^{4J} \frac{1}{|V|} \sum_{h,k} D_{hk} \leq \frac{1}{1-\g} J^2e^{4J},
$$
which completes the proof of Proposition \ref{teo}
\cvd

\br{powerlow}
The control of the total variation distance between $\pi^G$ and $\pi^{PCA}$ by
\be{pl}
\sup_{\d\in[0,\d_0]}\frac{1}{|V|}Var_{\p^{PCA}}\left[\sum_{i\in V} \frac{\phi_i}{1+\d\phi_i}\right],
\ee
may be useful even if the above quantity diverges as $|V| \rightarrow +\infty$. For instance, for the one dimensional model with
\[
J_{i,j} := \frac{J_1}{|i-j|^2},
\]
it is known the existence of an intermediate phase for which the spin-spin correlations decay as $|i-j|^{-2+\e}$, for some $\e \in (0,2]$ (see \cite{IN}). Assuming that a similar decay hold for the correlations of the $\varphi_i$'s, the quantity in (\ref{pl}) is expected to behave as $|V|^{\e}$ as $|V| \uparrow +\infty$. Thus, for Theorem \ref{teo} to hold, we need
\[
\lim_{|V| \uparrow +\infty} \d^2 |V|^{1+\e} = 0
\]
that, for $\e < 1$, still allow $\d |V| \ra +\infty$, i.e. a large number of spin updates per step.
\er

%%%%%%

%%%%%%%%%%%%%%%%%%SEC3%%%%%%%%%%%%%
%%%%%%%%%%%%%%%%%%
\section{The mean field Ising  model}
\label{CW}

As an example we discuss the performances of the PCA dynamics for the {\em mean-field Ising model}, or Curie-Weiss model. In this section some 
computations are given only at a heuristic level. The rigorous treatment would be straightforward but rather lengthy.

\subsection{Distance between  $\p^{PCA}$ and  $\p^G$ and comparison of phase diagrams}

We consider now the following mean field hamiltonian on $\cX:=\{-1,+1\}^{\{1,...,n\}}$
\be{HCW}
H_{CW}(\s)=-{J\over 2n}\sum_{i, j}\s_i\s_j
\ee
and the corrisponding pair hamiltonian
\be{HCWpair}
H_{CW}(\s,\s')=-{J\over 2n}\sum_{i,j}\s_i\s'_j+q\sum_i(1-\s_i\s'_i)
\ee
By using the  definitions
$
m=m(\s)={1\over n}\sum_i\s_i, 
$
we can study the mean field model  in the standard way. Indeed
we have immediately
$$
H_{CW}(\s)=-{J\over 2}nm^2
$$
so that
$$
\p^G(m):=\sum_{\s:\, m(\s)=m}\p^G(\s)=\frac{e^{nF(m)+o(n)}}{Z^G}
$$
with
$$F(m)= {J\over 2}m^2+I({1+m\over 2}),\qquad I(x):=-x\ln x-(1-x)\ln (1-x).$$
The small remainder $o(n)$ is such that $\frac{o(n)}{n}\to 0$ uniformly for $|m|<1-\e$
for any fixed $\e$. The contribution of the magnetizations close to $\pm 1$ can be shown
to be negligible. 
Moreover

$$
Z^G=e^{ n F(m^*)+o(n)} 
$$
where
 $m^*:=\arg\max F(m)$ satisfies the standard condition for the Curie-Weiss model:
$$Jm^*= {1\over 2}\ln{1+m^*\over 1-m^*}$$
obtaining $m^*=0$ for $J<1$ while for $J>1$ the solutions $m^*_+=-m^*_-$ can be obtained graphically.

With an immediate computation we get for the function $f$ defined in (\ref{f}):
\be{pif}
f(m)=\Big( 1+\d e^{-Jm} \Big)^{n\frac{m+1}{2}}\Big( 1+\d e^{Jm} \Big)^{n\frac{1-m}{2}}=: e^{n g(m,\d)}
\ee
with 
$$
g(m,\d):={a+b\over 2}+m{a-b\over 2}
$$
with 
$$a=a(m,\d):=\ln(1+\d e^{-Jm}),\qquad b=b(m,\d):=\ln(1+\d e^{Jm})$$
The function $g(m)=g(m,\d)$ as a function of $m$, is a $C^\infty$ function with the following properties:
$$g(0)=\ln(1+\d),\qquad g'(0)=0, \qquad g''(0) = -\d J(1-J)$$
\begin{equation} \label{g'}
g'(m) = -\d mJ \cosh(Jm) + \d(J-1) \sinh(Jm) + o(\d)
\end{equation}
$$g(-m)=g(m).$$

By defining $\bar m:=\arg\max F(m)+\d g(m)$ we have
that the measure $\p^{PCA}$ is concentrated on the configurations
with magnetization $\bar m$. 

In the case $J<1$ it is immediate to verify that
$\bar m=m^*=0$. Recalling that
\[
F(m) = F(0) - \frac{1-J}{2} m^2 + o(m^2),
\]
the probability measures $\p^{G}$ and $\p^{PCA}$
 can be estimated by  Gaussian distributions centered in $0$ with variances $[n(1-J)]^{-1}$ and $[n(1-J-\d J(1-J))]^{-1}$ respectively. 
Therefore we can compute explicitly 
\[
\frac{f(m)}{\pi^G(f)} = e^{n(g(m) - g(0))}(1+o_n(1)),
\]
yielding, for small $\d$ and large $n$,
\[
{\p^G(f^2)\over (\p^G(f))^2}-1= \sum_m\left( {e^{nF(m)}\over Z^G}{f(m)\over \p^G(f)}\right)\left[{f(m)\over \p^G(f)}
-1\right] 
\]
\[
 = \sum_m \pi^{PCA}(m) \left[{f(m)\over \p^G(f)} 
-1\right]  \simeq \sum_m \pi^{PCA}(m)\left[e^{-n\d J(1-J)m^2/2} -1\right] 
\]
\[
\simeq \sqrt{\frac{n(1-J-\d J(1-J))}{2\pi}} \int dm \, e^{-n(1-J-\d J(1-J)) m^2/2} \left[ e^{-n\d J(1-J)m^2/2} -1\right] \simeq \frac{J \d}{2}.
\]

Hence we obtain in this case  a  result stronger than Theorem \ref{teo} since we do not need the hypothesis
$\d^2 n\to 0$, being
 enough that $\d\to 0$ as $n\to\infty$.
 
 In the {low temperature  case $J>1$ }, for any finite $J$ (temperature
strictly positive), again we obtain the convergence result when $\d^2n\to 0$, if
the system is restricted to a single phase. 
From the dynamical point of view the restriction to a single phase can be obtained
simply by a reflecting barrier, following for instance  \cite{LLP}. Consider the dynamics on $\cX_+:=\{\s: m(\s)\ge 0\}$
obtained by generating a candidate move $\s '\in\cX$ according to (\ref{pss}) with the pair hamiltonian (\ref{HCWpair}),
and accepting it as a new state if $\s '\in \cX_+$ and adopting $-\s '$ as new state if $\s '\not\in \cX_+$,
obtaining in this way for $\s,\s'\in\cX^+$ a transition probability
$$
P^+(\s,\s')= P^{PCA}(\s,\s')+P^{PCA}(\s,-\s')
$$
This dynamics on $\cX_+$ is reversible w.r.t. the  invariant mesaure restricted to $\cX_+$: for each $\s,\s'\in{\cX}_+$
$$
\p^{PCA}|_{\cX_+}(\s)P^+(\s,\s')={\p^{PCA}(\s)\over \p^{PCA}({\cX_+})}\Big[P^{PCA}(\s,\s')+P^{PCA}(\s,-\s')\Big]$$
$$
={\p^{PCA}(\s')\over \p^{PCA}({\cX_+})}\Big[P^{PCA}(\s',\s)+P^{PCA}(-\s',\s)\Big]=\p^{PCA}|_{\cX_+}(\s')P^+(\s',\s),
$$
where we used the fact that $\pi^{PCA}(\s') = \pi^{PCA}(-\s')$.
Moreover with the same argument used in the high temperature case, we obtain 
$|m^*-\bar m|={\cal O}(\d)$ and 
$$
{\p^G(f^2)\over (\p^G(f))^2} -1 =\sum_m\left( {e^{nF(m)}\over Z^G}{f(m)\over \p^G(f)}\right)\left[{f(m)\over \p^G(f)} -1\right] \simeq \sum_m \pi^{PCA}(m) \left[ e^{n(g(m) - g(m^*))} -1 \right]$$

$$
\sim e^{n[g(\bar m)-g(m^*)]} -1 \sim e^{n g'(m^*)(\bar m - m^*)} -1  = O(\d^2 n),
$$
where we used the fact that $g'(m) = O(\d)$ (see (\ref{g'})).

\subsection{Convergence to equilibrium}

We give an estimate of the mixing time of the PCA for high temperature
($J<1$) using a classical coupling argument. We first recall that the transition probabilities of the PCA obey the following identity:
\be{tp}
P(\s_i'|\s) = \frac{e^{\s_i'(J m(\s) + q \s_i)}}{2 \cosh((J m(\s) + q \s_i)}.
\ee
Given two configurations $\s^+$ and $\s^-$ we will write $\s^+\succeq\s^-$
if $\s_i^+\geq\s_i^-\quad \forall\ i\in V$. 
We will define
a coupling  of the transition probabilities $P(\s^{'+}|\s^+)$ and $P(\s^{'-}|\s^-)$
in the following way: we extract for each site $i\in V$ an independent random
variable $u_i$ uniformly distributed in $[0,1]$. Then, using {\it the same} random variable for
both realizations of our PCA we say that $\s'^{\pm}_i=-1$ if $u_i\le P(\s'^{\pm}_i=-1|\s^{\pm})$
and $\s'^{\pm}_i=+1$ otherwise. It is immediate to see that this updating rule gives a marginal 
distribution for both $\s'^+$ and $\s'^-$ which is the original distribution of our PCA chain, and 
that if $\s^+\succeq\s^-$ then also $\s'^+\succeq\s'^-$. If we now denote with 
$n_{\rm diff}=\frac{n}{2}(m^+-m^-)$ the number 
of sites $i\in V$ such that $\s_i^+>\s_i^-$, and with $n'_{\rm diff}$ the number 
of sites $i\in V$ such that $\s'^+_i>\s'^-_i$, using (\ref{tp}) and the coupling rule we have that
\begin{eqnarray} \label{diff}
E(n'_{\rm diff}|\s^{\pm}) & = & \frac{n}{2}(m^+-m^-)\left(1-\frac{e^{-Jm^+-2q}}{1+e^{-Jm^+-2q}}-1 + 
\frac{1}{1+e^{Jm^--2q}}\right) \nonumber \\ & +
& \frac{n}{2}(1+m^-)\left(\frac{e^{-Jm^--2q}}{1+e^{-Jm^--2q}}-
\frac{e^{-Jm^+-2q}}{1+e^{Jm^+-2q}}\right) \\ & + & 
\frac{n}{2}(1-m^+)\left(\frac{1}{1+e^{Jm^--2q}}-
\frac{1}{1+e^{Jm^+-2q}}\right). \nonumber
\end{eqnarray}
This can be written, up to the first order in $e^{-2q}=\d$, as
\be {diff2}
E(n'_{\rm diff}|\s^{\pm})=n_{\rm diff}-n\d(m^+\cosh Jm^+-m^-\cosh Jm^--\sinh Jm^++\sinh Jm^-)
+O(\d^2)
\ee
Using now the fact that 
$$\sinh Jm^\pm\le Jm^\pm\cosh Jm^\pm$$ 
we have that

\be {diff3}
E(n'_{\rm diff}|\s^{\pm})\le n_{\rm diff}-n\d\frac{1-J}{J}(Jm^+\cosh Jm^+-\sinh Jm^-)
+O(\d^2)
\ee
and by the inequality
$$Jm^+\cosh Jm^+-\sinh Jm^-\ge J(m^+-m^-)$$ 
we obtain finally
\be {diff4}
E(n'_{\rm diff}|\s^{\pm})\le n_{\rm diff}(1-2\d(1-J)+O(\d^2))
\ee
Consider now the coupling applied to two copies of the PCA starting with spins all $+1$ and $-1$ respectively. 
When $J<1$, (\ref{diff4}) shows that $E(n_{\rm diff}(t))$ contracts exponentially in $t$. Denoting by $\tau_c$ the {\em coalescing} time of the two chains, we have
\[
P(\tau_c > t) \leq P(n_{\rm diff}(t) \neq 0) \leq E(n_{\rm diff}(t)) \leq 2n \left[1-2\d(1-J)\right]^t
\]
By monotonicity, the total variation distance from equilibrium at time $t$   of a PCA chain starting from an arbitrary configuration is bounded above by $P(\tau_c > t)$. This implies that the mixing time $T_{mix}$ is of order $\frac{\log n}{(1-J)\d}$.

\vspace{1cm}

\noindent {\bf Acknowledgments:} 
The authors wish to thank Pierre Picco for suggesting reference \cite{Fo82}, Errico Presutti and Alain Messager for fruitful discussions. \\
This work has been partially supported by the PRIN projects ``Campi aleatori, percolazione ed evoluzione stocastica di sistemi con molte componenti'', ``Dal microscopico al macroscopico: analisi di strutture complesse e applicazioni'', and by the GDRE 224 GREFI-MEFI-CNRS-INdAM.

\vglue15.truecm

\end{document}